\documentclass[twocolumn,amsmath,amssymb,superscriptaddress,floatfix]{revtex4}
\usepackage{color}
\usepackage{epsfig}
\usepackage{epstopdf}
\usepackage{grffile}
\usepackage{bm}
\usepackage{sidecap}
\usepackage{mathptmx}
\usepackage{txfonts}
\usepackage[colorlinks,citecolor=blue,linkcolor=blue,urlcolor=blue]{hyperref}
\usepackage[usenames,dvipsnames,svgnames]{xcolor}
\usepackage{array}
\usepackage{float}

\usepackage{lipsum}
\usepackage{graphicx}
\usepackage{dcolumn}
\usepackage{bm}
\usepackage{tabularx}
\usepackage{tabu}
\usepackage{multirow}
\usepackage{array}
\usepackage{booktabs}
\usepackage{tabularx} 
\usepackage{ulem}
\renewcommand{\bm }{\mathbf}
\newcommand{\PL}{\textnormal{PL}}
\renewcommand{\eqref}[1]{Eq. (\ref{#1})}
\newcommand{\tableref}[1]{Table \ref{#1}}

\begin{document}
\title{Dark-exciton based strain sensing in transition metal dichalcogenides}

\author{Maja Feierabend}
\affiliation{Chalmers University of Technology, Department of Physics, 412 96 Gothenburg, Sweden}
\author{Zahra Khatibi}
\affiliation{Chalmers University of Technology, Department of Physics, 412 96 Gothenburg, Sweden}
\affiliation{Iran University of Science and Technology, Department of Physics,  Narmak, 16846-13114, Tehran, Iran}
\author{Gunnar Bergh{\"a}user}
\affiliation{Chalmers University of Technology, Department of Physics, 412 96 Gothenburg, Sweden}
\author{Ermin Malic}
\affiliation{Chalmers University of Technology, Department of Physics, 412 96 Gothenburg, Sweden}

\begin{abstract}
The trend towards ever smaller high-performance devices in modern technology requires novel materials with new functionalities. The recent emergence of atomically thin two-dimensional (2D) materials has opened up possibilities for the design of ultra-thin and flexible nanoelectronic devices. As truly 2D materials, they exhibit an optimal surface-to-volume ratio, which results in an extremely high sensitivity to external changes. This makes these materials optimal candidates for sensing applications. Here, we exploit the remarkably diverse exciton landscape in monolayer transition metal dichalcogenides to propose a novel dark-exciton-based concept for ultra sensitive strain sensors. We demonstrate that the dark-bright-exciton separation can be controlled by strain, which has a crucial impact on the activation of dark excitonic states. This results in a pronounced intensity change of dark excitons in photoluminescence spectra, when  only 0.05 $\%$ strain is applied. The predicted  extremely high optical gauge factors of up to 8000 are promising for the design of optical strain sensors.
\end{abstract}

\maketitle

\section{Introduction}

Currently, the most common type of strain sensors consists of a metallic foil pattern, which becomes deformed in presence of strain resulting in a change of the electrical resistance \cite{zhou2008flexible,wang2006piezoelectric, sirohi2000fundamental}. In the case of small strain values, piezo-resistors are more appropriate \cite{hicks2014piezoelectric, kon2007piezoresistive, youn2008measurement}, since they have a larger gauge factor that is typically in the range of 10-20. 
The latter is defined as the ratio between the relative change in the electrical resistance and the applied mechanical strain. However, they are also more sensitive to temperature changes and are more fragile than foil-based sensors. The recently discovered class of atomically thin nanomaterials including graphene and monolayer transition metal dichalcogenides (TMDs) have the potential for the design of novel strain sensing devices \cite{boland2014sensitive,tsai2015flexible}.
Consisting of a single layer of atoms, they have an optimal surface-to-volume ratio resulting in an extremely high sensitivity to external stimuli. 
First graphene-based strain sensors show large gauge factors in the range of 35-500 \cite{casiraghi2018inkjet, li2012stretchable, zhao2012ultra}. Values reported so far for $\rm{MoS_2}$ are in the range 40-140 \cite{park2016mos2, tsai2015flexible}, while the recently studied $\rm{PtSe_2}$ monolayers reveal a gauge factor of 85 \cite{wagner2018highly}. However, the mentioned sensors are all based on a change in resistance due to strain, which makes them also sensitive to change in temperature or humidity.

\begin{figure}[t!]
  \begin{center}
\includegraphics[width=\linewidth]{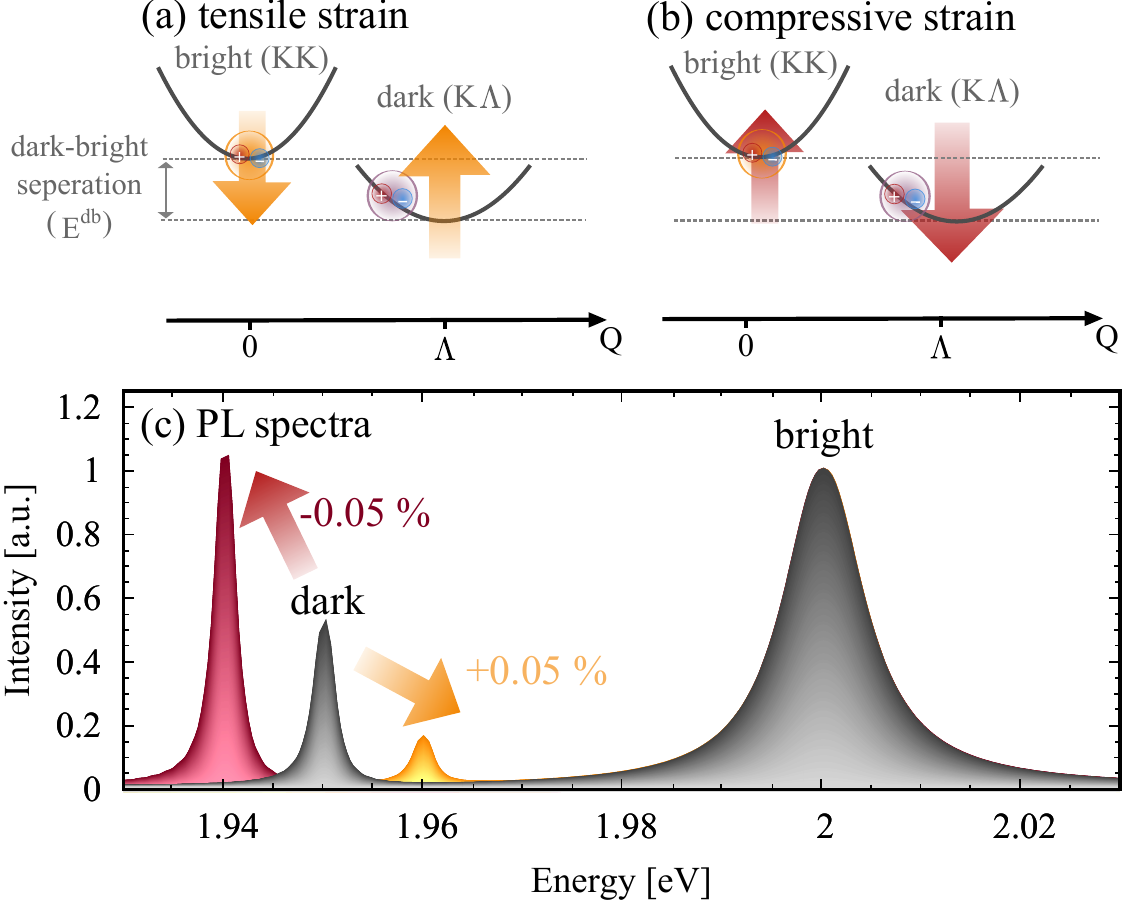} 
\end{center}
    \caption{\textbf{Impact of strain on exciton dispersion and PL.} (a) Tensile and (b) compressive strain shifts bright and momentum-forbidden dark exciton states in opposite direction. As a result, the spectral dark-bright separation $E^{\text{db}}$ is strongly sensitive to strain.   
    (c) Photoluminescence (PL) spectra for monolayer WS$_2$ including the bright resonance at 2.0 eV  and the phonon-, disorder-, or molecule-activated dark resonance at 1.95 eV in the unstrained case (grey line). Applying  0.05$\%$ tensile (compressive)
strain leads to a blue(red)-shift and a strong decrease (increase) in PL intensity of the dark exciton peak. This results in an optical gauge factor of 800 for the tensile and 8000 for the compressive strain. Note that the position of the bright exciton has been kept constant to focus on relative changes. }
   \label{schema}
\end{figure}

In this work, we investigate another possible mechanism for sensing of strain  based on dark excitons in 2D materials.
TMD monolayers exhibit a remarkably rich exciton physics due to the strong Coulomb interaction resulting in the formation of bright and spin- and momentum-forbidden dark excitonic states \cite{malic2018dark,selig2018dark,hsu2017evidence, deilmann2017dark}.
The latter cannot be directly accessed by light due to the required spin flip or large momentum-transfer. In previous studies, it was shown that that the spin-forbidden states consisting of Coulomb-bound electrons and holes with opposite spin can be brightened up either by applying strong in-plane magnetic fields \cite{stier2016probing,molas2017brightening} or via  coupling to surface plasmons \cite{zhou2017probing}. Momentum-forbidden states, where the Coulomb-bound electrons and holes are located in different valleys (K, $\Lambda$) in the momentum space, can be activated through interaction with phonons \cite{christiansen2017phonon} or high-dipole molecules \cite{maja_sensor, feierabend2018molecule} providing the required momentum to reach these states. Here, we show that the spectral separation of bright and momentum-forbidden dark excitonic states can be sensitively controlled by application of strain (Fig. \ref{schema}(a)-(b)). This has a huge impact on the photoluminescence (PL) intensity of activated dark states. A tensile strain of only 0.05\% reduces the PL of the dark state by one order of magnitude (Fig. \ref{schema}(c)). This extremely high sensitivity suggests the possibility to design novel atomically thin dark-exciton-based strain sensors. 
\section{Theoretical approach}
To investigate strain-induced changes in the optical response of monolayer TMDs, we combine density functional theory (DFT) calculations providing access to the electronic band structure with density matrix formalism allowing us to calculate the excitonic binding energies and the excitonic photoluminescence.
As a first step, we obtain the electronic structure of the unstrained 
monolayer tungsten disulfide (WS$_2$) that we investigate throughout the work as an exemplary TMD material.
To this end, we relax the lattice structure using ultrasoft pseudopotentials with LDA exchange-correlation functionals alongside the plane waves implemented in the \textsc{Quantum Espresso} package \cite{giannozzi2009quantum}. The momentum space is sampled with $18\times 18 \times 1$ Monkhorst--pack mesh and the kinetic energy cutoff is set to $60$ Ry. The total forces on each atom after relaxation is less than $0.001 \frac{\text{Ry}}{\text{a.u.}}$. 
To avoid the possible interaction of adjacent layers, we use a vacuum space of approximately 20~\AA~perpendicular to $\rm{WS_2}$ layer.
For the unstrained case, we obtain a lattice constant of $a_0=0.315$ nm, which is in good agreement with experimental values and previous DFT calculations \cite{andor}.  In the next step, we model the strained $\rm{WS_2}$ monolayer via the change of the lattice constant relative to the strain with $a^s_0=a_0(1+\varepsilon_s)$. Here, $\varepsilon_s$ is the applied strain, while $a^s_0$ and $a_0$ denote the strained and unstrained lattice constant, respectively. Then, we relax the new lattice, while keeping the lattice vectors fixed at their strained value. Therefore, only atomic positions can relax with regard to the strained vectors. Note that we use the same energy mesh cutoff and k-grid for the strained $\rm{WS_2}$. Finally, we utilize non-self-consistent field calculation to obtain the electronic dispersion for both relaxed unstrained and strained $\rm{WS_2}$ monolayer.

\begin{figure}[t!]
  \begin{center}
\includegraphics[width=\linewidth]{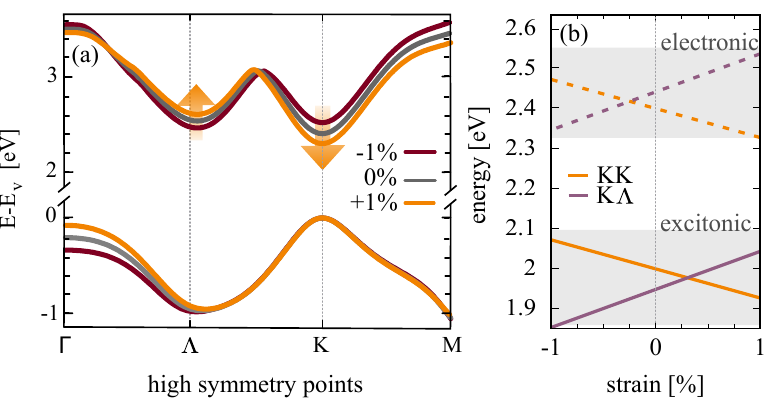} 
\end{center}
    \caption{\textbf{Impact of strain on the electronic band structure.} (a) DFT calculations of the electronic dispersion of  the WS$_{2}$ monolayer along the high-symmetry points in presence of tensile ($+1\%$) and compressive ($-1\%$) biaxial strain. To focus on relative changes, we use the valence band maximum at the K point as the reference. We find that the conduction band at the K point shifts down (up) by 150 meV at $\pm1\%$ of tensile (compressive) strain. In contrast, the $\Lambda$ point shows the qualitatively opposite behavior resulting in an up(down)-shifts by 40  meV for tensile (compressive) strain. Note that the absolute value of the band gap was adjusted to experimentally observed values \cite{niehues2018strain}. (b) Strain-induced change of the energy of the bright KK and the dark K$\Lambda$ exciton. While the first  shows a decrease with strain (orange), the latter exhibits a positive slope resulting in a cross-over at a certain strain value. The behavior can be mostly traced back to the strain-dependence of the electronic K and $\Lambda$ valley. The dashed lines show the direct and indirect electronic gaps $E^{KK}$ and $E^{K\Lambda}$. }
   \label{bands}
\end{figure}

To investigate strain-induced changes in the optical response of monolayer TMDs, we use the density matrix formalism providing access to excitonic binding energies, wavefunctions, and photoluminescence \cite{gunnar_prb}.
We extract material-specific values as DFT input parameters for the solution of the Wannier equation, cf.  \tableref{shifts}. This includes the curvature $m_{v (c)}^{K/\Lambda}$ of the valence (v) and the conduction (c) band at the high-symmetry K and $\Lambda$ points as well as the direct electronic gap  $E^{KK}$ between the conduction and the valence band with same spin at the K point. The indirect electronic gap $E^{K\Lambda}$ can be expressed as
$
E^{K\Lambda}= E^{KK}+ \Delta E^{K\Lambda}
$
with $ \Delta E_{K\Lambda}$ as the energy difference from the conduction band minimum at the K and the $\Lambda$ point. Surprisingly, we find that strain has a qualitatively different impact on the K and the $\Lambda$ valley: the conduction band minimum at the K point shifts down (up) for tensile (compressive) strain compared to the fixed valence band maximum, whereas the conduction band minimum at the $\Lambda$ point shows the opposite behavior, cf. Fig. \ref{bands}(a). 
Note that actually all band extrema show a down (up) shift for tensile (compressive) strain, however the conduction band at the K valley shifts stronger than that at the $\Lambda$ valley resulting in an opposite shift relative to the valence band. This can be understood as a consequence of the different orbital composition of the conduction band: while at the K point it consists mainly of $d_{z^2}$ orbitals of the tungsten atoms and $p_x+p_y$ orbitals of the sulfur atoms, at  the $\Lambda$ valley $d_{xy}, d_{x^2+y^2} $ and $d_{z^2}$ orbitals of the tungsten atoms play the major role \cite{chang2013orbital,yue2012mechanical}. When applying strain to the system, the distance between the atoms is changed and hence the overlap of the corresponding orbitals, leading to different strain rates in the shifts.

\begin{table}[b!]
\centering
\caption{\textbf{Electronic and excitonic quantities under strain.} Strain induced changes in the electronic dispersion around the high symmetry $K$ and $\Lambda$ points including the effective mass of the conduction band $m_c$ [$m_0$], the direct electronic gap $E^{KK}$ [eV], and the energetic difference between direct and the indirect electronic band gap $\Delta E^{K\Lambda}$ [eV].  Interestingly, the direct and the indirect gaps shift in opposite direction under strain. The spectral position of the bright exciton $E^{KK}_{exc}$ [eV] and the corresponding dark-bright splitting $E^{\text{db}}=E^{K\Lambda}_{\text{exc}}-E^{KK}_{\text{exc}}$ [eV]  are obtained by evaluating the Wannier equation. }\label{shifts}
\begin{tabular}{c >{\hspace{.07cm}} 
c>{\hspace{.07cm}}
c>{\hspace{.07cm}}
c>{\hspace{.07cm}} c>{\hspace{.07cm}} c>{\hspace{.07cm}} c} 
\hline
 strain&  ${\rm m_c^{K}}$ &  ${\rm m_c^{\Lambda}}$ &  $E^{KK}$ &  $\Delta E^{K\Lambda}$ &$E^{KK}_{\text{exc}}$ & $E^{\text{db}}$ \\ 
\hline
0  		&  0.26 			&  0.5  		&  2.4 			&0.050  				& 2.0   	 		&0.050\\ 
$\pm 1 \%$ & $\pm0.013 $& $\pm0.0 $ 	&  $\mp0.14 $ 	&$\pm0.19 $		& $\mp0.12 $ 	&$\pm0.19 $\\
\hline
\end{tabular}
\end{table}

The strain-induced changes in the electronic dispersion at high-symmetry points are also shown in \tableref{shifts}.
For  $\pm 1\%$ tensile (compressive) strain, the WS$_{2}$ monolayer  is stretched (compressed) and the lattice constant is changed to $a_0=0.318$ nm ($a_0=0.312$ nm). 
For the unstrained case (grey line in Fig. \ref{bands}), we find the direct band gap at the K point. However, the minimum in the conduction band at the $\Lambda$ point is close in energy with $\Delta E^{K\Lambda}=50$ meV. This distance is crucially affected by strain. When applying 1 $\%$ tensile strain (yellow curve), the conduction band minimum at the K point shifts down in energy by -140 meV, while the conduction band minimum at the $\Lambda$ point shows an opposite behaviour and shifts up by 50 meV relatively to the position of the valence band maximum at the K point. For compressive strain (red curve), we observe the opposite behavior. As a result, we find a change of the indirect electronic gap $E^{K\Lambda}$   by $\pm$ 190 meV/$\%$, where $+$ corresponds to the tensile and  $-$ to the compressive strain. 

Regarding the curvature of the bands and hence the effective masses, we find only a relatively small change due to strain with $m_{c}^{K(\Lambda)}$ being modified by 2 $\%$. The obtained values are in good agreement with previous DFT studies \cite{yue2012mechanical, shi2013quasiparticle}. Note that we focus in this work on biaxial strain, which means that the lattice is deformed uniformly in all directions and thus strain can be modeled by changing the lattice constant. For uniaxial strain, our DFT calculation reveal a similar behaviour regarding changes in energy shifts and effective masses. 
However, it is worth noting that in case of the uniaxial strain the broken lattice symmetry is transferred to the orbital shape which can lead to a softening of valley-dependent optical selection rules. However, this effect is rather small \cite{feierabend2017impact} and thus 1$\%$ biaxial strain roughly corresponds to 2 $\%$ uniaxial strain for the investigated range of relatively small strain values. 

So far, we have discussed the impact of strain on the electronic properties of an exemplary TMD material. However, it is crucial to study the change of excitonic properties, which determine the optical response of these materials. Exploiting  this DFT input, we can calculate the excitonic dispersion by solving the Wannier equation
\begin{equation}\label{wannier}
 \frac{\hbar^2 q^{2}}{2m^{\mu}}  \varphi_{\bf q}^{\mu} 
 - \sum_{\bf k} V_{\text{exc}}(\bf k)  \varphi_{\bf {q-k} }^{\mu}=\varepsilon_b^{\mu}\varphi_{\bf {q}}^{\mu}.
\end{equation}
This is an eigenvalue equation for excitons with the excitonic eigen functions $\varphi_{\bf q}^{\mu}$  and the excitonic binding energies $\varepsilon_b^{\mu}$, where ${\mu}$ is the exciton index including in particular the $\rm{KK}$ and the K$\Lambda$ exciton.
Furthermore, $V_{\text{exc}}$ describes the Coulomb-induced formation of excitons within the Keldysh potential \cite{Keldysh1978, gunnar_prb}. The exciton binding energy depends on the the reduced exciton mass $m^{\mu}=\frac{m_{v}^{K}\, m_{c}^{\mu}}{m_{v}^{K}+ m_{c}^{\mu}}$, where the conduction and valence band masses enter from our DFT calculations. For WS$_2$ on SiO$_2$ substrate, we obtain a binding energy of $\varepsilon^{KK}=400$ meV for the bright KK exciton and $\varepsilon^{K\Lambda}=500$ meV for the dark K$\Lambda$ exciton. The latter is larger due to the heavier mass of the $\Lambda$ valley. Since strain has only a minor influence on the effective masses (\tableref{shifts}), the excitonic binding energy is only slightly affected by strain \cite{feierabend2017impact}.

The spectral position of the exciton can  be expressed as 
$
E_{\text{exc}}^{KK} = E^{KK} -\varepsilon_b^{KK}$ and $E_{\text{exc}}^{K\Lambda} = E^{K\Lambda} -\varepsilon_b^{K\Lambda}$.
Note that the larger exciton binding energy of the K$\Lambda$ state exceeds the spectral differences in the electronic band structure between the K and the $\Lambda$ valley. As a result, the dark K$\Lambda$ exciton  becomes  lower in energy than the bright KK state in tungsten-based TMDs \cite{malic2018dark,selig2018dark}. 
We can write $E_{\text{exc}}^{K\Lambda}=E_{\text{exc}}^{KK} - E^{\text{db}}$, where $E^{\text{db}}$ corresponds to the spectral dark-bright exciton separation, cf. Fig. \ref{schema}(a). For the unstrained case, $E^{\text{db}} = 50 $ meV in the WS$_2$ monolayer suggesting that this material is an indirect-gap semiconductor in the excitonic picture.

To calculate the optical response of the system, we use the framework of the density matrix formalism in second quantization. The key quantity here is the photon number $ n_q = \langle c^{\dagger}_{\bf q}c_{ \bf q} \rangle$ with photon creation/annihilation operators $ c^{\dagger}_{\bf q}$. The steady-state photoluminescence $\PL(\omega_q)$  is given by the rate of emitted photons
$
\PL(\omega_q) \propto \omega_q \frac{\partial}{\partial t} \langle c^{\dagger}_{\bf q}c_{ \bf q}\rangle,
$ 
where  $\omega_q$ is the photon frequency.
Exploiting the Heisenberg equation of motion, we obtain for the dynamics of the photon number $
\dot { n_q }= [n_q, H]
$. The many-particle Hamilton operator $H$  includes the light-matter interaction describing the optical excitation of the system and the Coulomb interaction giving rise to the formation of strongly bound excitons. Furthermore, we include a mechanism for the activation of the momentum-forbidden K$\Lambda$ excitons. This can be driven either by phonons, molecules or actual disorder/impurity sites in the TMD lattice. 
In a recent PL study, indications for phonon-assisted radiative recombination  pathways to momentum-forbidden excitons have been observed \cite{lindlau2017identifying,lindlau2017role}.
In previous theoretical work \cite{feierabend2018molecule, maja_sensor}, we have discussed in detail the interaction of excitons with high-dipole molecules providing the required momentum to reach the dark K$\Lambda$ excitons. 

The dynamics of  the photon number $n_q$ is driven by photon-assisted polarization \cite{thranhardt2000quantum} 
  $S^{vc_{\mu}}_{\bf{k_1}\bf{k_2}} =  \langle c^{\dagger}_{\bf q}
a^{\dagger v}_{\bf{k_1}} a^{c_\mu}_{\bf{k_2}}
\rangle $ with electron annihilation (creation) $a^{(\dagger)}$ and photon annihilation (creation) $c^{(\dagger)}$ operators. Solving the corresponding set of equations we determine the steady-state photoluminescence on microscopic footing:
\begin{equation}\label{PLana}
 I(\omega) \propto
 \text{Im} \bigg(
 \frac{|M_{\omega}^{\sigma K}|^2}{\Delta E_{\omega}^{KK} - 
\frac{|G^{K\Lambda}|^2}{\Delta E_{\omega}^{K\Lambda}}}
\left[
N_{\bf 0}^{KK} (1-\alpha)  
 + 
\alpha N_{\bf 0}^{K\Lambda} 
 \right]
\bigg).
\end{equation}
Here,  $ M_{\omega}^{\sigma K}$  is the optical matrix element describing the optical excitation of the K valley with the circularly polarized light $\sigma$. It determines the optical oscillator strength of excitonic resonances. 
The position of the latter in the PL spectrum is given by
 $\Delta E_{\omega}^{\mu} = E_{exc}^{\mu} - \hbar \omega - i\gamma^{\mu}$ in the denominator. 
Finally, we have introduced an abbreviation $\alpha= \frac{|G^{K\Lambda}|^2}{(\Delta E_{\omega}^{\Lambda})(E^{\text{db}}-i\gamma^{K\Lambda})}$, where the matrix element $G^{K\Lambda}$ is given by the overlap of the excitonic wave functions 
$
G_{\bf{Qk}}^{\mu\nu}= \sum_{\bf{q}} 
(
\varphi_{\bf {q}}^{\mu*} g^{cc}_{\bf{q}_\alpha, \bf{q}_\alpha + \bf{k}} \varphi_{\bf{q+\beta k}}^{\nu} 
 - 
\varphi_{\bf {q}}^{\mu*} g^{vv}_{\bf{q}_\beta - \bf{k}, \bf{q}_\beta} \varphi_{\bf {q-\alpha k}}^{\nu})
$
with $\bf{q}_\alpha=\bf{q}-\alpha\bf{Q}$ and $\bf{q}_\beta=\bf{q}+\beta\bf{Q}$. The exact form of the coupling element $g$ depends on the actual interaction that is necessary to activate the momentum-forbidden dark K$\Lambda$ excitonic state. This can i.a. driven by disorder, phonons, molecules or any other momentum generating mechanism. The particular mechanism is not important for the current study.

We find that the PL is strongly influenced by exciton occupations $N_{\bf 0}^{KK}, N_{\bf 0}^{K\Lambda}$ with a vanishing center-of-mass momentum $\bm Q \approx \bm 0$. Only these states are located in the light cone and can emit light through radiative recombination. After the process of phonon-induced exciton thermalization occurring on a time scale of 1 ps \cite{selig2018dark}, the exciton occupations can be described by a Bose distribution. 
 Figure \ref{schema}(c) shows the calculated PL spectrum for unstrained WS$_2$ monolayer at the exemplary temperature of 77 K (grey line). The peak at 2.0 eV corresponds to the bright KK exciton, while the energetically lower peak at 1.95 eV stems from the dark K$\Lambda$ exciton. In the following section, we discuss in detail how these excitonic signatures in the PL spectrum change in presence of tensile and compressive strain.  

\section{Results}

Combining the DFT input for the strain-induced change of the electronic dispersion and the solution of the Wannier equation providing access to the excitonic binding energies, we calculate the exact spectral position of  dark and bright excitonic states, cf. Fig. \ref{bands}(b).
While the energy of the bright KK exciton decreases with tensile strain, the momentum-forbidden dark K$\Lambda$ exciton shifts to higher energies. 
The strain-induced change can be fitted by a linear function with the slope $-$ 140 meV/$\%$ for the bright and $+$ 40 meV/$\%$ for the dark state. As a result, the spectral dark-bright separation $E^{\text{db}}$ changes sign at 0.3 $\%$ of tensile strain. 
This strongly influences the efficiency of the activation of dark K$\Lambda$ excitonic state and has a direct impact on its position and intensity in the PL spectrum. 
For comparison, the dashed lines in Fig. \ref{bands}(b) visualize the direct and indirect electronic gaps $E^{KK}$ and $E^{K\Lambda}$ (without taking into account excitonic effects). We observe that the indirect gap lies energetically above the direct gap in the unstrained case. However, the strain behaviour is qualitatively the same reflecting the fact that the exciton binding energy is not very sensitive to strain \cite{feierabend2017impact}. As a result, it is the opposite strain-dependence of the direct and indirect electronic gaps that determines the dark-bright-exciton separation and thus also the spectral features and their strain sensitivity. 

\begin{figure}[t!]
  \begin{center}
\includegraphics[width=\linewidth]{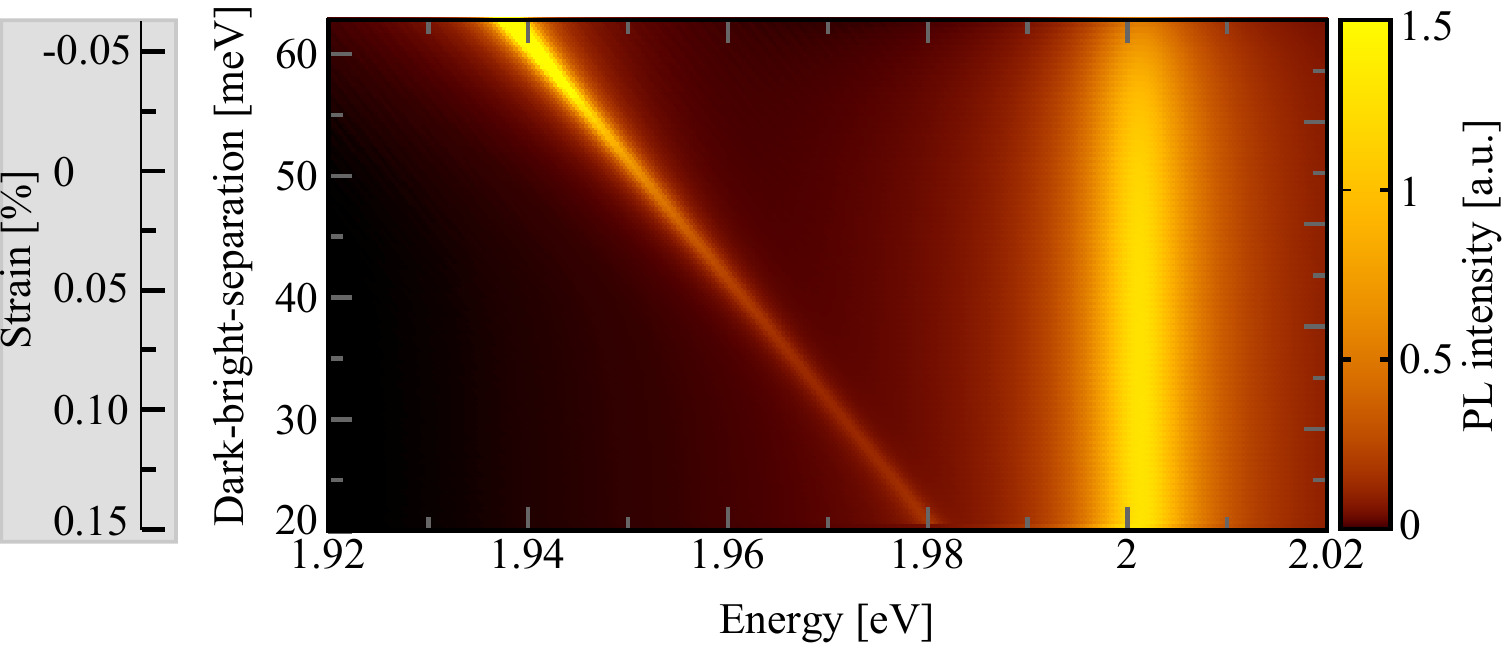} 
\end{center}
    \caption{\textbf{Impact of strain on dark-bright-exciton separation and PL intensity.} Spectrally resolved PL of the WS$_{2}$ monolayer as a function of the  dark-bright exciton separation $E^{\text{db}}$, which can be directly translated to a certain strain value. According to our calculations, the unstrained WS$_{2}$  exhibits a dark-bright separation of 50 meV. Applying compressive strain (negative strain values), the energetic separation
becomes larger, while for tensile strain (positive value) strain, it becomes smaller.   We observe a pronounced change in the PL intensity of the dark K$\Lambda$, as it comes closer to the bright state. Note that the position and intensity of the bright KK exciton has been fixed to focus on relative strain-induced changes. 
}
   \label{EKQ}
\end{figure}

Figure \ref{EKQ} reveals the PL intensity of the WS$_2$ monolayer as a function of energy and the dark-bright splitting (and strain) at the exemplary temperature of 77K. 
The constant broad peak at 2 eV stems from the bright exciton, while the shifting peak at the lower energy is ascribed to the dark K$\Lambda$ exciton that has been activated either via phonons, disorder or molecules.  As the dark-bright separation $E^{\text{db}}$ is the key quantity determining the strain-induced change in the optical response, we have fixed the bright exciton resonance at 2.0 eV for better visualization.
The unstrained case corresponds to an energetic difference  $ E^{\text{db}} \approx 50$ meV corresponding to the PL in Fig. \ref{schema}(c). Interestingly, the smallest amounts of strain lead to pronounced changes in the optical response. This is due to the high sensitivity of the position and intensity of the dark exciton resonance to  $ E^{\text{db}}$ that itself is highly sensitive to the applied strain. We predict a strain-induced change rate of $\pm$ 190 meV/$\%$ for $ E^{\text{db}}$ (\tableref{shifts}).

The strong change in the PL intensity of the dark exciton as the tensile strain increases and $ E^{\text{db}}$ decreases can be traced back to the change in the exciton population of the involved KK and K$\Lambda$ excitons. After the system is thermalized, the exciton population is given by the Bose distribution, where the energetic position of each excitonic state explicitly enters.  If the material is tensile strained, the K$\Lambda$ exciton shifts up in energy, while the KK exciton shifts down (Fig. \ref{schema}(a)) resulting in a reduced dark-bright separation (Fig. \ref{bands}). As a result, it becomes unfavourable for excitons to occupy the dark state. As a direct consequence, the intensity of the dark exciton is considerably reduced and the dark exciton can even be fully deactivated under certain strain values. 

On the other hand, applying compressive strain shifts the dark exciton even lower in energy, resulting in a larger dark-bright-exciton separation $E^{\text{db}}$ and a higher occupation of the dark exciton state after thermalization. Hence, we predict the intensity of the dark exciton to significantly increase for compressive strain. Our calculations show that considerable changes in the optical response appear, when applying only 0.05 $\%$ of strain, cf. Fig. \ref{EKQ}. This suggest that smallest amounts of strain could be detected by measuring a simple PL spectrum. Note that as the TMD material is usually placed on a substrate, the strain is actually transferred from the substrate to the TMD layer. Depending on the transfer rate and the Poisson ratio of the substrate, the actual transferred strain to the TMD monolayer can be externally controlled. As a result, the extremely sensitive strain range depending on the TMD material and the used substrate can be shifted to the technologically desired values.

\begin{figure}[t!]
  \begin{center}
\includegraphics[width=\linewidth]{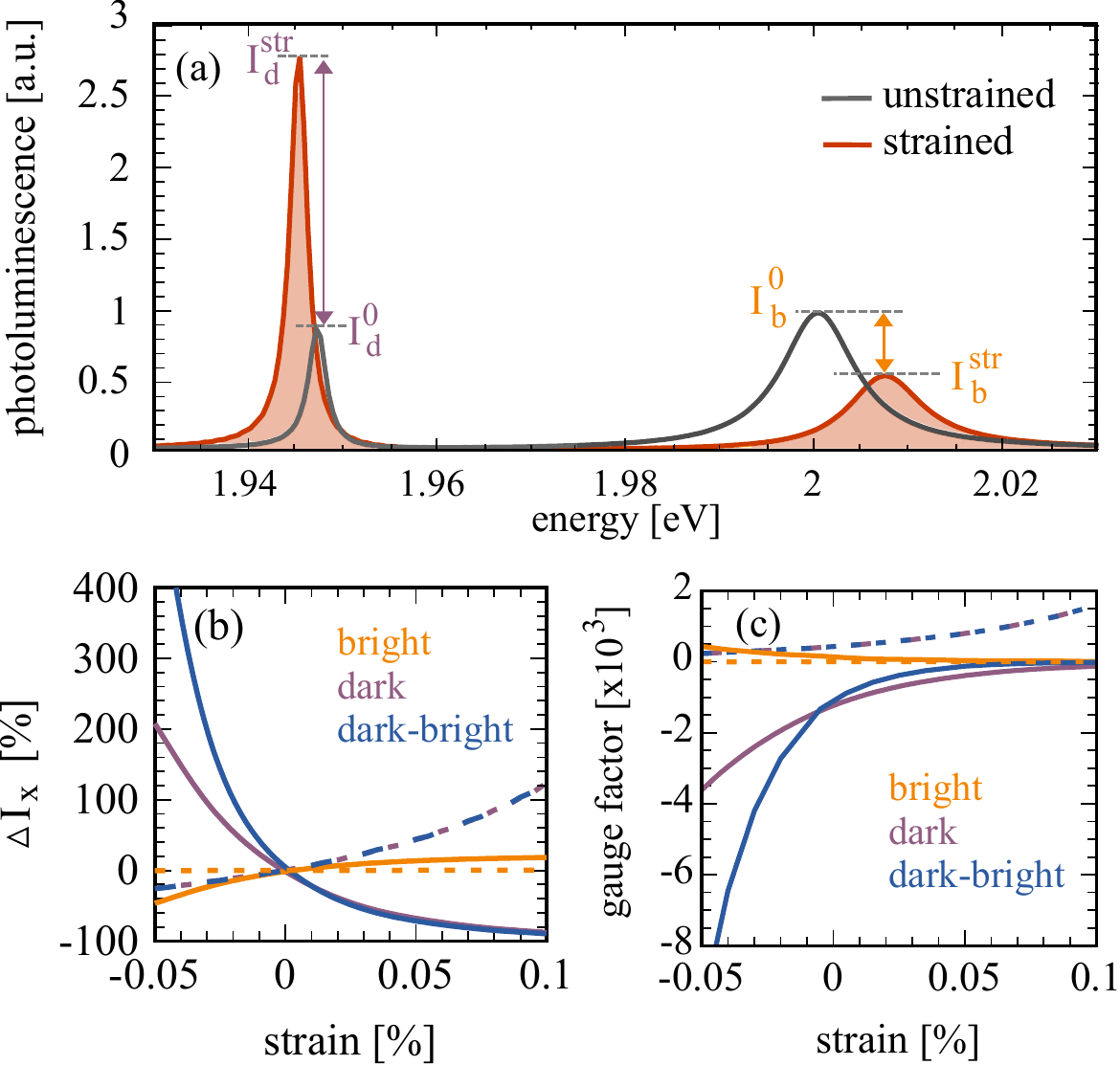} 
\end{center}
    \caption{\textbf{Optical gauge factors.} (a) PL spectrum of $\rm{WS_2}$ for the unstrained and the unstrained (-0.05\%) case at 77K. Note that here the strain-induced shift and intensity change of the bright exciton is also shown. The extracted (b) intensity changes and (c) optical gauge factors $g_x$  , where x= b, d denoting the bright KK and the dark K$\Lambda$ exciton, respectively. The gauge factors are based on the intensity ratio  between the unstrained  ($I_x^0$) and the strained case ($I_x^{\rm{str}}$) for the dark exciton $g_d$ (purple) and the bright exciton $g_b$ (yellow), cf. Eq. (\ref{gaugi}). We also show the corresponding gauge factor based on the change of the dark-bright exciton ratio in presence of strain (blue). Dashed lines show  the behaviour at room temperature.}
   \label{gauge}
\end{figure}

An important quantity for strain sensors is the gauge factor corresponding to the ratio between the relative strain-induced change in a certain property of the material and the actually applied mechanical strain.  For our work, the optical gauge factor, defined as the ratio of the relative change in PL intensity and the strain,  is a measure for the sensitivity of the dark-exciton-based strain sensor 
\begin{equation}
    \rm{g_{x}} = 
    \frac{\Delta I_x [\%]}
    {\varepsilon_s [\%]}.
    \label{gaugi}
\end{equation}
with the applied strain $\varepsilon_s [\%]$ and the intensity ratio $\Delta I_x=\frac{I_x^{\text{str}}- I_x^{0}}{I_x^0}$, where $x$ describes the bright KK or the dark K$\Lambda$ exciton. Figure \ref{gauge} shows the PL spectrum for unstrained  and strained $\rm{WS_2}$ as well as strain-dependent intensity ratios and optical gauge factors.  In addition to the bright (dark) gauge factor $g_b$ ($g_d$) that are determined by the strain-induced intensity changes of the bright (dark) KK (K$\Lambda$) exciton, we also consider the dark-bright gauge factor $g_{db}$ that is based on the change of the intensity ratio between the dark and the bright exciton, i.e. $g_{db}=\frac{I_d^{\text{str}}/I_b^{\text{str}}}{I_d^{0}/I_b^{0}}-1$. 
Figures \ref{gauge}(b) and (c) show the corresponding changes in the relative intensity ratio $\Delta I_x$ and the gauge factors $g_x$ as a function of strain at 77 K (solid lines) and 300K (dashed lines). In the case of a  linear strain dependence, the gauge factors simply correspond to  the slope of the corresponding lines. However, since the intensity changes are clearly non-linear, the gauge factors vary with strain and can be understood as the derivative of the relative intensity ratio. 

Our calculations reveal  that one can obtain remarkable high gauge factors for the compressive strain at 77K. The dark-bright gauge factor $g_{db}$ can reach values of up to 8000 for 0.05 $\%$ of compressive strain (blue line). The dark gauge factor $g_d$ (purple line) goes up to 4000, while the bright gauge factor $g_b$ (yellow curve) is in the range of 500. 
These values exceed reported values for 2D materials by an order of magnitude (35-500 for graphene \cite{casiraghi2018inkjet, li2012stretchable, zhao2012ultra}, 40-140 for $\rm{MoS_2}$ \cite{park2016mos2, tsai2015flexible} , 85 for $\rm{PtSe2}$ \cite{wagner2018highly}). 
At room temperature (dashed lines), the gauge factors become lower in general, however they are still high for $g_d$ and $g_{db}$ reaching values up to 1500, whereas $g_b$ considerably decreases to roughly 10. As a result, at room temperature the dark gauge factors are two orders of magnitude higher then the bright gauge factor, opening up possible applications for ultrasensitive sensing of strain.

To understand the high sensitivity of the dark exciton towards strain, we recapture its photoluminescence properties, cf.  \eqref{PLana}. The PL intensity is determined by exciton occupations $N^{x}$ that are given by the Bose distribution in equilibrium, where in particular $N^{K\Lambda} \propto N^{KK} \text{exp}[{-E^{\text{db}}(k_bT)^{-1}}]$ with the spectral bright-dark separation $E^{\text{db}}$. As a result, we find for the strained intensity $I^{\text{str}} \propto \text{exp}[{-E^{\text{db}}_{\text{str}}(k_bT)^{-1}}]$, i.e. there is an exponential decrease of the PL intensity with strain, as observed in Fig. \ref{gauge}(b). For compressive (tensile) strain, $E^{\text{db}}$ increases (decreases) and hence the dark state is stronger (weaker) occupied resulting in an enhanced (reduced) PL intensity. At the same time, the bright exciton shows the opposite behavior, since its occupation increases or decreases depending on the relative position of the dark state, cf. Figs. \ref{gauge} (b) and (c). As a result, we find remarkably high dark-bright gauge factor $g_{db}$ in presence of compressive strain, where the strain-induced changes of the dark and the bright exciton add up, i.e.  the PL intensity of the dark exciton increases and at the same time the intensity of the bright peak decreases. For tensile strain, the behavior is opposite, since  the dark peak becomes smaller, while  the bright is enhanced resulting in   $g_{db} < g_d$, cf. purple and blue lines in Fig. \ref{gauge}(c). 

For tensile strain values larger than 0.1\%, we find that all gauge factors go asymptotically towards zero at 77 K. In presence of tensile strain, dark and bright excitons come closer together (Fig. \ref{EKQ}) and the corresponding peaks start to spectrally overlap in the PL spectrum. Moreover, the dark exciton becomes less occupied, as it is higher in energy and hence disappears at some point. 
Note that for compressive strain, a saturation is also reached at higher strain values reflecting the saturation of the dark exciton state. 
At room temperature (dashed lines in Figs. \ref{gauge}(b), (c)) the behaviour is different due to the much larger phonon-induced broadening of excitonic resonances. 
Here, the spectral overlap of the dark and the bright peak becomes more important than the exciton occupation. 
The closer the states become, the higher is the PL intensity of the dark peak. As a result, in presence of tensile strain larger dark and dark-bright gauge factors are predicted.

So far, we have studied and discussed the strain dependence in a $\rm{WS_2}$ monolayer as an exemplary TMD material.  We expect a comparable behaviour for $\rm{WSe_2}$, as it shows a similar electronic and excitonic dispersion, in particular with respect to the relative position of the dark K$\Lambda$ exciton \cite{malic2018dark}. The case of $\rm{MoS_2}$ is a bit more complicated, since here the energetically lowest $\Gamma$K exciton also needs to be taken into account. Since this exciton requires a rather large momentum transfer to be activated, we expect $\rm{MoS_2}$ to be less appropriate for strain sensing. Finally, $\rm{MoSe_2}$ is the only direct-gap semiconductor in the excitonic picture \cite{malic2018dark}, where all indirect excitons lie above the bright state. It still could be used for strain sensing, as strain can move dark and bright states closer to each other and enable the activation of dark excitons in PL spectra.

\section{Conclusion}
In conclusion, we have investigated the strain-induced control of the optical response in a $\rm{WS_2}$ monolayer as an exemplary material for transition metal dichalcogenides. In particular, we have focused on the  change of the photoluminescence intensity of bright  and activated dark excitonic states. We find that the changes in the electronic band structure play the crucial role, since they determine the spectral
separation of dark and bright excitonic states. We predict a remarkable sensitivity to strain resulting in optical gauge factors in the range of up to 8000. The obtained insights are the first step towards ultrasensitive strain sensing based on dark excitons in atomically thin materials.

\section*{Acknowledgement}
This project has received funding from the European
Union's Horizon 2020 research and innovation programme
under grant agreement No 696656. Furthermore,
we acknowledge financial support from the Swedish Research
Council (VR).

\bibliography{refererences_maja2} 

\begin{thebibliography}{35}
\expandafter\ifx\csname natexlab\endcsname\relax\def\natexlab#1{#1}\fi
\expandafter\ifx\csname bibnamefont\endcsname\relax
  \def\bibnamefont#1{#1}\fi
\expandafter\ifx\csname bibfnamefont\endcsname\relax
  \def\bibfnamefont#1{#1}\fi
\expandafter\ifx\csname citenamefont\endcsname\relax
  \def\citenamefont#1{#1}\fi
\expandafter\ifx\csname url\endcsname\relax
  \def\url#1{\texttt{#1}}\fi
\expandafter\ifx\csname urlprefix\endcsname\relax\def\urlprefix{URL }\fi
\providecommand{\bibinfo}[2]{#2}
\providecommand{\eprint}[2][]{\url{#2}}

\bibitem[{\citenamefont{Zhou et~al.}(2008)\citenamefont{Zhou, Gu, Fei, Mai,
  Gao, Yang, Bao, and Wang}}]{zhou2008flexible}
\bibinfo{author}{\bibfnamefont{J.}~\bibnamefont{Zhou}},
  \bibinfo{author}{\bibfnamefont{Y.}~\bibnamefont{Gu}},
  \bibinfo{author}{\bibfnamefont{P.}~\bibnamefont{Fei}},
  \bibinfo{author}{\bibfnamefont{W.}~\bibnamefont{Mai}},
  \bibinfo{author}{\bibfnamefont{Y.}~\bibnamefont{Gao}},
  \bibinfo{author}{\bibfnamefont{R.}~\bibnamefont{Yang}},
  \bibinfo{author}{\bibfnamefont{G.}~\bibnamefont{Bao}}, \bibnamefont{and}
  \bibinfo{author}{\bibfnamefont{Z.~L.} \bibnamefont{Wang}},
  \bibinfo{journal}{Nano letters} \textbf{\bibinfo{volume}{8}},
  \bibinfo{pages}{3035} (\bibinfo{year}{2008}).

\bibitem[{\citenamefont{Wang et~al.}(2006)\citenamefont{Wang, Zhou, Song, Liu,
  Xu, and Wang}}]{wang2006piezoelectric}
\bibinfo{author}{\bibfnamefont{X.}~\bibnamefont{Wang}},
  \bibinfo{author}{\bibfnamefont{J.}~\bibnamefont{Zhou}},
  \bibinfo{author}{\bibfnamefont{J.}~\bibnamefont{Song}},
  \bibinfo{author}{\bibfnamefont{J.}~\bibnamefont{Liu}},
  \bibinfo{author}{\bibfnamefont{N.}~\bibnamefont{Xu}}, \bibnamefont{and}
  \bibinfo{author}{\bibfnamefont{Z.~L.} \bibnamefont{Wang}},
  \bibinfo{journal}{Nano letters} \textbf{\bibinfo{volume}{6}},
  \bibinfo{pages}{2768} (\bibinfo{year}{2006}).

\bibitem[{\citenamefont{Sirohi and Chopra}(2000)}]{sirohi2000fundamental}
\bibinfo{author}{\bibfnamefont{J.}~\bibnamefont{Sirohi}} \bibnamefont{and}
  \bibinfo{author}{\bibfnamefont{I.}~\bibnamefont{Chopra}},
  \bibinfo{journal}{Journal of intelligent material systems and structures}
  \textbf{\bibinfo{volume}{11}}, \bibinfo{pages}{246} (\bibinfo{year}{2000}).

\bibitem[{\citenamefont{Hicks et~al.}(2014)\citenamefont{Hicks, Barber, Edkins,
  Brodsky, and Mackenzie}}]{hicks2014piezoelectric}
\bibinfo{author}{\bibfnamefont{C.~W.} \bibnamefont{Hicks}},
  \bibinfo{author}{\bibfnamefont{M.~E.} \bibnamefont{Barber}},
  \bibinfo{author}{\bibfnamefont{S.~D.} \bibnamefont{Edkins}},
  \bibinfo{author}{\bibfnamefont{D.~O.} \bibnamefont{Brodsky}},
  \bibnamefont{and} \bibinfo{author}{\bibfnamefont{A.~P.}
  \bibnamefont{Mackenzie}}, \bibinfo{journal}{Review of Scientific Instruments}
  \textbf{\bibinfo{volume}{85}}, \bibinfo{pages}{065003}
  (\bibinfo{year}{2014}).

\bibitem[{kon()}]{kon2007piezoresistive}
 (????).

\bibitem[{\citenamefont{Youn et~al.}(2008)\citenamefont{Youn, Choo, and
  Kim}}]{youn2008measurement}
\bibinfo{author}{\bibfnamefont{J.-U.} \bibnamefont{Youn}},
  \bibinfo{author}{\bibfnamefont{Y.-W.} \bibnamefont{Choo}}, \bibnamefont{and}
  \bibinfo{author}{\bibfnamefont{D.-S.} \bibnamefont{Kim}},
  \bibinfo{journal}{Canadian Geotechnical Journal}
  \textbf{\bibinfo{volume}{45}}, \bibinfo{pages}{1426} (\bibinfo{year}{2008}).

\bibitem[{\citenamefont{Boland et~al.}(2014)\citenamefont{Boland, Khan, Backes,
  O'Neill, McCauley, Duane, Shanker, Liu, Jurewicz, Dalton
  et~al.}}]{boland2014sensitive}
\bibinfo{author}{\bibfnamefont{C.~S.} \bibnamefont{Boland}},
  \bibinfo{author}{\bibfnamefont{U.}~\bibnamefont{Khan}},
  \bibinfo{author}{\bibfnamefont{C.}~\bibnamefont{Backes}},
  \bibinfo{author}{\bibfnamefont{A.}~\bibnamefont{O'Neill}},
  \bibinfo{author}{\bibfnamefont{J.}~\bibnamefont{McCauley}},
  \bibinfo{author}{\bibfnamefont{S.}~\bibnamefont{Duane}},
  \bibinfo{author}{\bibfnamefont{R.}~\bibnamefont{Shanker}},
  \bibinfo{author}{\bibfnamefont{Y.}~\bibnamefont{Liu}},
  \bibinfo{author}{\bibfnamefont{I.}~\bibnamefont{Jurewicz}},
  \bibinfo{author}{\bibfnamefont{A.~B.} \bibnamefont{Dalton}},
  \bibnamefont{et~al.}, \bibinfo{journal}{ACS nano}
  \textbf{\bibinfo{volume}{8}}, \bibinfo{pages}{8819} (\bibinfo{year}{2014}).

\bibitem[{\citenamefont{Tsai et~al.}(2015)\citenamefont{Tsai, Tarasov, Hesabi,
  Taghinejad, Campbell, Joiner, Adibi, and Vogel}}]{tsai2015flexible}
\bibinfo{author}{\bibfnamefont{M.-Y.} \bibnamefont{Tsai}},
  \bibinfo{author}{\bibfnamefont{A.}~\bibnamefont{Tarasov}},
  \bibinfo{author}{\bibfnamefont{Z.~R.} \bibnamefont{Hesabi}},
  \bibinfo{author}{\bibfnamefont{H.}~\bibnamefont{Taghinejad}},
  \bibinfo{author}{\bibfnamefont{P.~M.} \bibnamefont{Campbell}},
  \bibinfo{author}{\bibfnamefont{C.~A.} \bibnamefont{Joiner}},
  \bibinfo{author}{\bibfnamefont{A.}~\bibnamefont{Adibi}}, \bibnamefont{and}
  \bibinfo{author}{\bibfnamefont{E.~M.} \bibnamefont{Vogel}},
  \bibinfo{journal}{ACS applied materials \& interfaces}
  \textbf{\bibinfo{volume}{7}}, \bibinfo{pages}{12850} (\bibinfo{year}{2015}).

\bibitem[{\citenamefont{Casiraghi et~al.}(2018)\citenamefont{Casiraghi,
  Macucci, Parvez, Worsley, Shin, Bronte, Borri, Paggi, and
  Fiori}}]{casiraghi2018inkjet}
\bibinfo{author}{\bibfnamefont{C.}~\bibnamefont{Casiraghi}},
  \bibinfo{author}{\bibfnamefont{M.}~\bibnamefont{Macucci}},
  \bibinfo{author}{\bibfnamefont{K.}~\bibnamefont{Parvez}},
  \bibinfo{author}{\bibfnamefont{R.}~\bibnamefont{Worsley}},
  \bibinfo{author}{\bibfnamefont{Y.}~\bibnamefont{Shin}},
  \bibinfo{author}{\bibfnamefont{F.}~\bibnamefont{Bronte}},
  \bibinfo{author}{\bibfnamefont{C.}~\bibnamefont{Borri}},
  \bibinfo{author}{\bibfnamefont{M.}~\bibnamefont{Paggi}}, \bibnamefont{and}
  \bibinfo{author}{\bibfnamefont{G.}~\bibnamefont{Fiori}},
  \bibinfo{journal}{Carbon} \textbf{\bibinfo{volume}{129}},
  \bibinfo{pages}{462} (\bibinfo{year}{2018}).

\bibitem[{\citenamefont{Li et~al.}(2012)\citenamefont{Li, Zhang, Yu, Wang, Wei,
  Wu, Cao, Li, Cheng, Zheng et~al.}}]{li2012stretchable}
\bibinfo{author}{\bibfnamefont{X.}~\bibnamefont{Li}},
  \bibinfo{author}{\bibfnamefont{R.}~\bibnamefont{Zhang}},
  \bibinfo{author}{\bibfnamefont{W.}~\bibnamefont{Yu}},
  \bibinfo{author}{\bibfnamefont{K.}~\bibnamefont{Wang}},
  \bibinfo{author}{\bibfnamefont{J.}~\bibnamefont{Wei}},
  \bibinfo{author}{\bibfnamefont{D.}~\bibnamefont{Wu}},
  \bibinfo{author}{\bibfnamefont{A.}~\bibnamefont{Cao}},
  \bibinfo{author}{\bibfnamefont{Z.}~\bibnamefont{Li}},
  \bibinfo{author}{\bibfnamefont{Y.}~\bibnamefont{Cheng}},
  \bibinfo{author}{\bibfnamefont{Q.}~\bibnamefont{Zheng}},
  \bibnamefont{et~al.}, \bibinfo{journal}{Scientific reports}
  \textbf{\bibinfo{volume}{2}}, \bibinfo{pages}{870} (\bibinfo{year}{2012}).

\bibitem[{\citenamefont{Zhao et~al.}(2012)\citenamefont{Zhao, He, Yang, Shi,
  Cheng, Yang, Xie, Wang, Shi, and Zhang}}]{zhao2012ultra}
\bibinfo{author}{\bibfnamefont{J.}~\bibnamefont{Zhao}},
  \bibinfo{author}{\bibfnamefont{C.}~\bibnamefont{He}},
  \bibinfo{author}{\bibfnamefont{R.}~\bibnamefont{Yang}},
  \bibinfo{author}{\bibfnamefont{Z.}~\bibnamefont{Shi}},
  \bibinfo{author}{\bibfnamefont{M.}~\bibnamefont{Cheng}},
  \bibinfo{author}{\bibfnamefont{W.}~\bibnamefont{Yang}},
  \bibinfo{author}{\bibfnamefont{G.}~\bibnamefont{Xie}},
  \bibinfo{author}{\bibfnamefont{D.}~\bibnamefont{Wang}},
  \bibinfo{author}{\bibfnamefont{D.}~\bibnamefont{Shi}}, \bibnamefont{and}
  \bibinfo{author}{\bibfnamefont{G.}~\bibnamefont{Zhang}},
  \bibinfo{journal}{Applied Physics Letters} \textbf{\bibinfo{volume}{101}},
  \bibinfo{pages}{063112} (\bibinfo{year}{2012}).

\bibitem[{\citenamefont{Park et~al.}(2016)\citenamefont{Park, Park, Chen, Park,
  Kim, and Ahn}}]{park2016mos2}
\bibinfo{author}{\bibfnamefont{M.}~\bibnamefont{Park}},
  \bibinfo{author}{\bibfnamefont{Y.~J.} \bibnamefont{Park}},
  \bibinfo{author}{\bibfnamefont{X.}~\bibnamefont{Chen}},
  \bibinfo{author}{\bibfnamefont{Y.-K.} \bibnamefont{Park}},
  \bibinfo{author}{\bibfnamefont{M.-S.} \bibnamefont{Kim}}, \bibnamefont{and}
  \bibinfo{author}{\bibfnamefont{J.-H.} \bibnamefont{Ahn}},
  \bibinfo{journal}{Advanced Materials} \textbf{\bibinfo{volume}{28}},
  \bibinfo{pages}{2556} (\bibinfo{year}{2016}).

\bibitem[{\citenamefont{Wagner et~al.}(2018)\citenamefont{Wagner, Yim, McEvoy,
  Kataria, Yokaribas, Kuc, Pindl, Fritzen, Heine, Duesberg
  et~al.}}]{wagner2018highly}
\bibinfo{author}{\bibfnamefont{S.}~\bibnamefont{Wagner}},
  \bibinfo{author}{\bibfnamefont{C.}~\bibnamefont{Yim}},
  \bibinfo{author}{\bibfnamefont{N.}~\bibnamefont{McEvoy}},
  \bibinfo{author}{\bibfnamefont{S.}~\bibnamefont{Kataria}},
  \bibinfo{author}{\bibfnamefont{V.}~\bibnamefont{Yokaribas}},
  \bibinfo{author}{\bibfnamefont{A.}~\bibnamefont{Kuc}},
  \bibinfo{author}{\bibfnamefont{S.}~\bibnamefont{Pindl}},
  \bibinfo{author}{\bibfnamefont{C.-P.} \bibnamefont{Fritzen}},
  \bibinfo{author}{\bibfnamefont{T.}~\bibnamefont{Heine}},
  \bibinfo{author}{\bibfnamefont{G.~S.} \bibnamefont{Duesberg}},
  \bibnamefont{et~al.}, \bibinfo{journal}{Nano letters}
  (\bibinfo{year}{2018}).

\bibitem[{\citenamefont{Malic et~al.}(2018)\citenamefont{Malic, Selig,
  Feierabend, Brem, Christiansen, Wendler, Knorr, and
  Bergh{\"a}user}}]{malic2018dark}
\bibinfo{author}{\bibfnamefont{E.}~\bibnamefont{Malic}},
  \bibinfo{author}{\bibfnamefont{M.}~\bibnamefont{Selig}},
  \bibinfo{author}{\bibfnamefont{M.}~\bibnamefont{Feierabend}},
  \bibinfo{author}{\bibfnamefont{S.}~\bibnamefont{Brem}},
  \bibinfo{author}{\bibfnamefont{D.}~\bibnamefont{Christiansen}},
  \bibinfo{author}{\bibfnamefont{F.}~\bibnamefont{Wendler}},
  \bibinfo{author}{\bibfnamefont{A.}~\bibnamefont{Knorr}}, \bibnamefont{and}
  \bibinfo{author}{\bibfnamefont{G.}~\bibnamefont{Bergh{\"a}user}},
  \bibinfo{journal}{Physical Review Materials} \textbf{\bibinfo{volume}{2}},
  \bibinfo{pages}{014002} (\bibinfo{year}{2018}).

\bibitem[{\citenamefont{Selig et~al.}(2018)\citenamefont{Selig, Bergh{\"a}user,
  Richter, Bratschitsch, Knorr, and Malic}}]{selig2018dark}
\bibinfo{author}{\bibfnamefont{M.}~\bibnamefont{Selig}},
  \bibinfo{author}{\bibfnamefont{G.}~\bibnamefont{Bergh{\"a}user}},
  \bibinfo{author}{\bibfnamefont{M.}~\bibnamefont{Richter}},
  \bibinfo{author}{\bibfnamefont{R.}~\bibnamefont{Bratschitsch}},
  \bibinfo{author}{\bibfnamefont{A.}~\bibnamefont{Knorr}}, \bibnamefont{and}
  \bibinfo{author}{\bibfnamefont{E.}~\bibnamefont{Malic}}, \bibinfo{journal}{2D
  Materials}  (\bibinfo{year}{2018}).

\bibitem[{\citenamefont{Hsu et~al.}(2017)\citenamefont{Hsu, Lu, Wang, Huang,
  Li, Chang, Chou, Juang, Jeng, Li et~al.}}]{hsu2017evidence}
\bibinfo{author}{\bibfnamefont{W.-T.} \bibnamefont{Hsu}},
  \bibinfo{author}{\bibfnamefont{L.-S.} \bibnamefont{Lu}},
  \bibinfo{author}{\bibfnamefont{D.}~\bibnamefont{Wang}},
  \bibinfo{author}{\bibfnamefont{J.-K.} \bibnamefont{Huang}},
  \bibinfo{author}{\bibfnamefont{M.-Y.} \bibnamefont{Li}},
  \bibinfo{author}{\bibfnamefont{T.-R.} \bibnamefont{Chang}},
  \bibinfo{author}{\bibfnamefont{Y.-C.} \bibnamefont{Chou}},
  \bibinfo{author}{\bibfnamefont{Z.-Y.} \bibnamefont{Juang}},
  \bibinfo{author}{\bibfnamefont{H.-T.} \bibnamefont{Jeng}},
  \bibinfo{author}{\bibfnamefont{L.-J.} \bibnamefont{Li}},
  \bibnamefont{et~al.}, \bibinfo{journal}{Nature communications}
  \textbf{\bibinfo{volume}{8}}, \bibinfo{pages}{929} (\bibinfo{year}{2017}).

\bibitem[{\citenamefont{Deilmann and Thygesen}(2017)}]{deilmann2017dark}
\bibinfo{author}{\bibfnamefont{T.}~\bibnamefont{Deilmann}} \bibnamefont{and}
  \bibinfo{author}{\bibfnamefont{K.~S.} \bibnamefont{Thygesen}},
  \bibinfo{journal}{Physical Review B} \textbf{\bibinfo{volume}{96}},
  \bibinfo{pages}{201113} (\bibinfo{year}{2017}).

\bibitem[{\citenamefont{Stier et~al.}(2016)\citenamefont{Stier, Wilson, Clark,
  Xu, and Crooker}}]{stier2016probing}
\bibinfo{author}{\bibfnamefont{A.~V.} \bibnamefont{Stier}},
  \bibinfo{author}{\bibfnamefont{N.~P.} \bibnamefont{Wilson}},
  \bibinfo{author}{\bibfnamefont{G.}~\bibnamefont{Clark}},
  \bibinfo{author}{\bibfnamefont{X.}~\bibnamefont{Xu}}, \bibnamefont{and}
  \bibinfo{author}{\bibfnamefont{S.~A.} \bibnamefont{Crooker}},
  \bibinfo{journal}{Nano letters} \textbf{\bibinfo{volume}{16}},
  \bibinfo{pages}{7054} (\bibinfo{year}{2016}).

\bibitem[{\citenamefont{Molas et~al.}(2017)\citenamefont{Molas, Faugeras,
  Slobodeniuk, Nogajewski, Bartos, Basko, and Potemski}}]{molas2017brightening}
\bibinfo{author}{\bibfnamefont{M.}~\bibnamefont{Molas}},
  \bibinfo{author}{\bibfnamefont{C.}~\bibnamefont{Faugeras}},
  \bibinfo{author}{\bibfnamefont{A.}~\bibnamefont{Slobodeniuk}},
  \bibinfo{author}{\bibfnamefont{K.}~\bibnamefont{Nogajewski}},
  \bibinfo{author}{\bibfnamefont{M.}~\bibnamefont{Bartos}},
  \bibinfo{author}{\bibfnamefont{D.}~\bibnamefont{Basko}}, \bibnamefont{and}
  \bibinfo{author}{\bibfnamefont{M.}~\bibnamefont{Potemski}},
  \bibinfo{journal}{2D Materials} \textbf{\bibinfo{volume}{4}},
  \bibinfo{pages}{021003} (\bibinfo{year}{2017}).

\bibitem[{\citenamefont{Zhou et~al.}(2017)\citenamefont{Zhou, Scuri, Wild,
  High, Dibos, Jauregui, Shu, De~Greve, Pistunova, Joe
  et~al.}}]{zhou2017probing}
\bibinfo{author}{\bibfnamefont{Y.}~\bibnamefont{Zhou}},
  \bibinfo{author}{\bibfnamefont{G.}~\bibnamefont{Scuri}},
  \bibinfo{author}{\bibfnamefont{D.~S.} \bibnamefont{Wild}},
  \bibinfo{author}{\bibfnamefont{A.~A.} \bibnamefont{High}},
  \bibinfo{author}{\bibfnamefont{A.}~\bibnamefont{Dibos}},
  \bibinfo{author}{\bibfnamefont{L.~A.} \bibnamefont{Jauregui}},
  \bibinfo{author}{\bibfnamefont{C.}~\bibnamefont{Shu}},
  \bibinfo{author}{\bibfnamefont{K.}~\bibnamefont{De~Greve}},
  \bibinfo{author}{\bibfnamefont{K.}~\bibnamefont{Pistunova}},
  \bibinfo{author}{\bibfnamefont{A.~Y.} \bibnamefont{Joe}},
  \bibnamefont{et~al.}, \bibinfo{journal}{Nature nanotechnology}
  \textbf{\bibinfo{volume}{12}}, \bibinfo{pages}{856} (\bibinfo{year}{2017}).

\bibitem[{\citenamefont{Christiansen et~al.}(2017)\citenamefont{Christiansen,
  Selig, Bergh{\"a}user, Schmidt, Niehues, Schneider, Arora, de~Vasconcellos,
  Bratschitsch, Malic et~al.}}]{christiansen2017phonon}
\bibinfo{author}{\bibfnamefont{D.}~\bibnamefont{Christiansen}},
  \bibinfo{author}{\bibfnamefont{M.}~\bibnamefont{Selig}},
  \bibinfo{author}{\bibfnamefont{G.}~\bibnamefont{Bergh{\"a}user}},
  \bibinfo{author}{\bibfnamefont{R.}~\bibnamefont{Schmidt}},
  \bibinfo{author}{\bibfnamefont{I.}~\bibnamefont{Niehues}},
  \bibinfo{author}{\bibfnamefont{R.}~\bibnamefont{Schneider}},
  \bibinfo{author}{\bibfnamefont{A.}~\bibnamefont{Arora}},
  \bibinfo{author}{\bibfnamefont{S.~M.} \bibnamefont{de~Vasconcellos}},
  \bibinfo{author}{\bibfnamefont{R.}~\bibnamefont{Bratschitsch}},
  \bibinfo{author}{\bibfnamefont{E.}~\bibnamefont{Malic}},
  \bibnamefont{et~al.}, \bibinfo{journal}{Physical review letters}
  \textbf{\bibinfo{volume}{119}}, \bibinfo{pages}{187402}
  (\bibinfo{year}{2017}).

\bibitem[{\citenamefont{Feierabend
  et~al.}(2017{\natexlab{a}})\citenamefont{Feierabend, Bergh{\"a}user, Knorr,
  and Malic}}]{maja_sensor}
\bibinfo{author}{\bibfnamefont{M.}~\bibnamefont{Feierabend}},
  \bibinfo{author}{\bibfnamefont{G.}~\bibnamefont{Bergh{\"a}user}},
  \bibinfo{author}{\bibfnamefont{A.}~\bibnamefont{Knorr}}, \bibnamefont{and}
  \bibinfo{author}{\bibfnamefont{E.}~\bibnamefont{Malic}},
  \bibinfo{journal}{Nature Communications} \textbf{\bibinfo{volume}{8}},
  \bibinfo{pages}{14776} (\bibinfo{year}{2017}{\natexlab{a}}).

\bibitem[{\citenamefont{Feierabend et~al.}(2018)\citenamefont{Feierabend,
  Bergh{\"a}user, Selig, Brem, Shegai, Eigler, and
  Malic}}]{feierabend2018molecule}
\bibinfo{author}{\bibfnamefont{M.}~\bibnamefont{Feierabend}},
  \bibinfo{author}{\bibfnamefont{G.}~\bibnamefont{Bergh{\"a}user}},
  \bibinfo{author}{\bibfnamefont{M.}~\bibnamefont{Selig}},
  \bibinfo{author}{\bibfnamefont{S.}~\bibnamefont{Brem}},
  \bibinfo{author}{\bibfnamefont{T.}~\bibnamefont{Shegai}},
  \bibinfo{author}{\bibfnamefont{S.}~\bibnamefont{Eigler}}, \bibnamefont{and}
  \bibinfo{author}{\bibfnamefont{E.}~\bibnamefont{Malic}},
  \bibinfo{journal}{Physical Review Materials} \textbf{\bibinfo{volume}{2}},
  \bibinfo{pages}{014004} (\bibinfo{year}{2018}).

\bibitem[{\citenamefont{Giannozzi et~al.}(2009)\citenamefont{Giannozzi, Baroni,
  Bonini, Calandra, Car, Cavazzoni, Ceresoli, Chiarotti, Cococcioni, Dabo
  et~al.}}]{giannozzi2009quantum}
\bibinfo{author}{\bibfnamefont{P.}~\bibnamefont{Giannozzi}},
  \bibinfo{author}{\bibfnamefont{S.}~\bibnamefont{Baroni}},
  \bibinfo{author}{\bibfnamefont{N.}~\bibnamefont{Bonini}},
  \bibinfo{author}{\bibfnamefont{M.}~\bibnamefont{Calandra}},
  \bibinfo{author}{\bibfnamefont{R.}~\bibnamefont{Car}},
  \bibinfo{author}{\bibfnamefont{C.}~\bibnamefont{Cavazzoni}},
  \bibinfo{author}{\bibfnamefont{D.}~\bibnamefont{Ceresoli}},
  \bibinfo{author}{\bibfnamefont{G.~L.} \bibnamefont{Chiarotti}},
  \bibinfo{author}{\bibfnamefont{M.}~\bibnamefont{Cococcioni}},
  \bibinfo{author}{\bibfnamefont{I.}~\bibnamefont{Dabo}}, \bibnamefont{et~al.},
  \bibinfo{journal}{Journal of physics: Condensed matter}
  \textbf{\bibinfo{volume}{21}}, \bibinfo{pages}{395502}
  (\bibinfo{year}{2009}).

\bibitem[{\citenamefont{Kormanyos et~al.}(2015)\citenamefont{Kormanyos,
  Burkard, Gmitra, Fabian, Zolyomi, Drummond, and Falko}}]{andor}
\bibinfo{author}{\bibfnamefont{A.}~\bibnamefont{Kormanyos}},
  \bibinfo{author}{\bibfnamefont{G.}~\bibnamefont{Burkard}},
  \bibinfo{author}{\bibfnamefont{M.}~\bibnamefont{Gmitra}},
  \bibinfo{author}{\bibfnamefont{J.}~\bibnamefont{Fabian}},
  \bibinfo{author}{\bibfnamefont{V.}~\bibnamefont{Zolyomi}},
  \bibinfo{author}{\bibfnamefont{N.~D.} \bibnamefont{Drummond}},
  \bibnamefont{and} \bibinfo{author}{\bibfnamefont{V.}~\bibnamefont{Falko}},
  \bibinfo{journal}{2D Materials} \textbf{\bibinfo{volume}{2}},
  \bibinfo{pages}{022001} (\bibinfo{year}{2015}).

\bibitem[{\citenamefont{Niehues et~al.}(2018)\citenamefont{Niehues, Schmidt,
  Dr{\"u}ppel, Marauhn, Christiansen, Selig, Bergh{\"a}user, Wigger, Schneider,
  Braasch et~al.}}]{niehues2018strain}
\bibinfo{author}{\bibfnamefont{I.}~\bibnamefont{Niehues}},
  \bibinfo{author}{\bibfnamefont{R.}~\bibnamefont{Schmidt}},
  \bibinfo{author}{\bibfnamefont{M.}~\bibnamefont{Dr{\"u}ppel}},
  \bibinfo{author}{\bibfnamefont{P.}~\bibnamefont{Marauhn}},
  \bibinfo{author}{\bibfnamefont{D.}~\bibnamefont{Christiansen}},
  \bibinfo{author}{\bibfnamefont{M.}~\bibnamefont{Selig}},
  \bibinfo{author}{\bibfnamefont{G.}~\bibnamefont{Bergh{\"a}user}},
  \bibinfo{author}{\bibfnamefont{D.}~\bibnamefont{Wigger}},
  \bibinfo{author}{\bibfnamefont{R.}~\bibnamefont{Schneider}},
  \bibinfo{author}{\bibfnamefont{L.}~\bibnamefont{Braasch}},
  \bibnamefont{et~al.}, \bibinfo{journal}{Nano letters}
  \textbf{\bibinfo{volume}{18}}, \bibinfo{pages}{1751} (\bibinfo{year}{2018}).

\bibitem[{\citenamefont{Bergh\"auser and Malic}(2014)}]{gunnar_prb}
\bibinfo{author}{\bibfnamefont{G.}~\bibnamefont{Bergh\"auser}}
  \bibnamefont{and} \bibinfo{author}{\bibfnamefont{E.}~\bibnamefont{Malic}},
  \bibinfo{journal}{Phys. Rev. B} \textbf{\bibinfo{volume}{89}},
  \bibinfo{pages}{125309} (\bibinfo{year}{2014}).

\bibitem[{\citenamefont{Chang et~al.}(2013)\citenamefont{Chang, Fan, Lin, and
  Kuo}}]{chang2013orbital}
\bibinfo{author}{\bibfnamefont{C.-H.} \bibnamefont{Chang}},
  \bibinfo{author}{\bibfnamefont{X.}~\bibnamefont{Fan}},
  \bibinfo{author}{\bibfnamefont{S.-H.} \bibnamefont{Lin}}, \bibnamefont{and}
  \bibinfo{author}{\bibfnamefont{J.-L.} \bibnamefont{Kuo}},
  \bibinfo{journal}{Physical Review B} \textbf{\bibinfo{volume}{88}},
  \bibinfo{pages}{195420} (\bibinfo{year}{2013}).

\bibitem[{\citenamefont{Yue et~al.}(2012)\citenamefont{Yue, Kang, Shao, Zhang,
  Chang, Wang, Qin, and Li}}]{yue2012mechanical}
\bibinfo{author}{\bibfnamefont{Q.}~\bibnamefont{Yue}},
  \bibinfo{author}{\bibfnamefont{J.}~\bibnamefont{Kang}},
  \bibinfo{author}{\bibfnamefont{Z.}~\bibnamefont{Shao}},
  \bibinfo{author}{\bibfnamefont{X.}~\bibnamefont{Zhang}},
  \bibinfo{author}{\bibfnamefont{S.}~\bibnamefont{Chang}},
  \bibinfo{author}{\bibfnamefont{G.}~\bibnamefont{Wang}},
  \bibinfo{author}{\bibfnamefont{S.}~\bibnamefont{Qin}}, \bibnamefont{and}
  \bibinfo{author}{\bibfnamefont{J.}~\bibnamefont{Li}}, \bibinfo{journal}{Phys.
  Lett. A} \textbf{\bibinfo{volume}{376}}, \bibinfo{pages}{1166}
  (\bibinfo{year}{2012}).

\bibitem[{\citenamefont{Shi et~al.}(2013)\citenamefont{Shi, Pan, Zhang, and
  Yakobson}}]{shi2013quasiparticle}
\bibinfo{author}{\bibfnamefont{H.}~\bibnamefont{Shi}},
  \bibinfo{author}{\bibfnamefont{H.}~\bibnamefont{Pan}},
  \bibinfo{author}{\bibfnamefont{Y.-W.} \bibnamefont{Zhang}}, \bibnamefont{and}
  \bibinfo{author}{\bibfnamefont{B.~I.} \bibnamefont{Yakobson}},
  \bibinfo{journal}{Phys. Rev. B} \textbf{\bibinfo{volume}{87}},
  \bibinfo{pages}{155304} (\bibinfo{year}{2013}).

\bibitem[{\citenamefont{Feierabend
  et~al.}(2017{\natexlab{b}})\citenamefont{Feierabend, Morlet, Bergh{\"a}user,
  and Malic}}]{feierabend2017impact}
\bibinfo{author}{\bibfnamefont{M.}~\bibnamefont{Feierabend}},
  \bibinfo{author}{\bibfnamefont{A.}~\bibnamefont{Morlet}},
  \bibinfo{author}{\bibfnamefont{G.}~\bibnamefont{Bergh{\"a}user}},
  \bibnamefont{and} \bibinfo{author}{\bibfnamefont{E.}~\bibnamefont{Malic}},
  \bibinfo{journal}{Physical Review B} \textbf{\bibinfo{volume}{96}},
  \bibinfo{pages}{045425} (\bibinfo{year}{2017}{\natexlab{b}}).

\bibitem[{\citenamefont{Keldysh}(1978)}]{Keldysh1978}
\bibinfo{author}{\bibnamefont{Keldysh}}, \bibinfo{journal}{JETP Lett.}
  \textbf{\bibinfo{volume}{29}}, \bibinfo{pages}{658} (\bibinfo{year}{1978}).

\bibitem[{\citenamefont{Lindlau
  et~al.}(2017{\natexlab{a}})\citenamefont{Lindlau, Robert, Funk, F{\"o}rste,
  F{\"o}rg, Colombier, Neumann, Courtade, Shree, Taniguchi
  et~al.}}]{lindlau2017identifying}
\bibinfo{author}{\bibfnamefont{J.}~\bibnamefont{Lindlau}},
  \bibinfo{author}{\bibfnamefont{C.}~\bibnamefont{Robert}},
  \bibinfo{author}{\bibfnamefont{V.}~\bibnamefont{Funk}},
  \bibinfo{author}{\bibfnamefont{J.}~\bibnamefont{F{\"o}rste}},
  \bibinfo{author}{\bibfnamefont{M.}~\bibnamefont{F{\"o}rg}},
  \bibinfo{author}{\bibfnamefont{L.}~\bibnamefont{Colombier}},
  \bibinfo{author}{\bibfnamefont{A.}~\bibnamefont{Neumann}},
  \bibinfo{author}{\bibfnamefont{E.}~\bibnamefont{Courtade}},
  \bibinfo{author}{\bibfnamefont{S.}~\bibnamefont{Shree}},
  \bibinfo{author}{\bibfnamefont{T.}~\bibnamefont{Taniguchi}},
  \bibnamefont{et~al.}, \bibinfo{journal}{arXiv preprint arXiv:1710.00988}
  (\bibinfo{year}{2017}{\natexlab{a}}).

\bibitem[{\citenamefont{Lindlau
  et~al.}(2017{\natexlab{b}})\citenamefont{Lindlau, Selig, Neumann, Colombier,
  Kim, Bergh{\"a}user, Wang, Malic, and H{\"o}gele}}]{lindlau2017role}
\bibinfo{author}{\bibfnamefont{J.}~\bibnamefont{Lindlau}},
  \bibinfo{author}{\bibfnamefont{M.}~\bibnamefont{Selig}},
  \bibinfo{author}{\bibfnamefont{A.}~\bibnamefont{Neumann}},
  \bibinfo{author}{\bibfnamefont{L.}~\bibnamefont{Colombier}},
  \bibinfo{author}{\bibfnamefont{J.}~\bibnamefont{Kim}},
  \bibinfo{author}{\bibfnamefont{G.}~\bibnamefont{Bergh{\"a}user}},
  \bibinfo{author}{\bibfnamefont{F.}~\bibnamefont{Wang}},
  \bibinfo{author}{\bibfnamefont{E.}~\bibnamefont{Malic}}, \bibnamefont{and}
  \bibinfo{author}{\bibfnamefont{A.}~\bibnamefont{H{\"o}gele}},
  \bibinfo{journal}{arXiv preprint arXiv:1710.00989}
  (\bibinfo{year}{2017}{\natexlab{b}}).

\bibitem[{\citenamefont{Thr{\"a}nhardt
  et~al.}(2000)\citenamefont{Thr{\"a}nhardt, Kuckenburg, Knorr, Meier, and
  Koch}}]{thranhardt2000quantum}
\bibinfo{author}{\bibfnamefont{A.}~\bibnamefont{Thr{\"a}nhardt}},
  \bibinfo{author}{\bibfnamefont{S.}~\bibnamefont{Kuckenburg}},
  \bibinfo{author}{\bibfnamefont{A.}~\bibnamefont{Knorr}},
  \bibinfo{author}{\bibfnamefont{T.}~\bibnamefont{Meier}}, \bibnamefont{and}
  \bibinfo{author}{\bibfnamefont{S.}~\bibnamefont{Koch}},
  \bibinfo{journal}{Physical Review B} \textbf{\bibinfo{volume}{62}},
  \bibinfo{pages}{2706} (\bibinfo{year}{2000}).

\end{thebibliography}
\end{document}